\documentclass[12pt]{iopart}
\usepackage{iopams}
\usepackage{graphics}

\def\empile#1\over#2{\mathrel{\mathop{\kern 0pt#1}\limits_{#2}}}

\newcommand{\slv}{\raise.15ex\hbox{$/$}\kern-.53em\hbox{$v$}}
\newcommand{\slF}{\raise.15ex\hbox{$/$}\kern-.53em\hbox{$F$}}
\newcommand{\slL}{\raise.15ex\hbox{$/$}\kern-.53em\hbox{$L$}}
\newcommand{\slP}{\raise.15ex\hbox{$/$}\kern-.53em\hbox{$P$}}
\newcommand{\slp}{\raise.15ex\hbox{$/$}\kern-.53em\hbox{$p$}}
\newcommand{\slq}{\raise.15ex\hbox{$/$}\kern-.53em\hbox{$q$}}
\newcommand{\slR}{\raise.15ex\hbox{$/$}\kern-.53em\hbox{$R$}}
\newcommand{\slQ}{\raise.15ex\hbox{$/$}\kern-.53em\hbox{$Q$}}
\newcommand{\slK}{\raise.15ex\hbox{$/$}\kern-.53em\hbox{$K$}}
\newcommand{\slk}{\raise.15ex\hbox{$/$}\kern-.53em\hbox{$k$}}
\newcommand{\slD}{\raise.15ex\hbox{$/$}\kern-.53em\hbox{$D$}}
\newcommand{\slC}{\raise.15ex\hbox{$/$}\kern-.53em\hbox{$C$}}
\newcommand{\slA}{\raise.15ex\hbox{$/$}\kern-.53em\hbox{$A$}}
\newcommand{\slSigma}{\raise.15ex\hbox{$/$}\kern-.53em\hbox{$\Sigma$}}
\newcommand{\slpartial}{\raise.15ex\hbox{$/$}\kern-.53em\hbox{$\partial$}}
\newcommand{\slcalP}{\raise.15ex\hbox{$/$}\kern-.63em\hbox{$\cal P$}}

\catcode`\@=11

\newcount\@tempcntc
\def\@citex[#1]#2{\if@filesw\immediate\write\@auxout{\string\citation{#2}}\fi
  \@tempcnta\z@\@tempcntb\m@ne\def\@citea{}\@cite{%
        \@for\@citeb:=#2\do%
    {\@ifundefined{b@\@citeb}%
        {\@citeo\@tempcntb\m@ne\@citea%
                \def\@citea{,\penalty\@m\ }{\bf ?}\@warning%
                {Citation `\@citeb' on page \thepage \space undefined}}%
        {\setbox\z@\hbox{\global\@tempcntc0\csname b@\@citeb\endcsname\relax}
     \ifnum\@tempcntc=\z@ \@citeo\@tempcntb\m@ne%
       \@citea\def\@citea{,\penalty\@m}%
       \hbox{\csname b@\@citeb\endcsname}%
     \else%
      \advance\@tempcntb\@ne%
      \ifnum\@tempcntb=\@tempcntc%
      \else\advance\@tempcntb\m@ne\@citeo%
      \@tempcnta\@tempcntc\@tempcntb\@tempcntc\fi\fi}}\@citeo}{#1}}%

\def\@citeo{\ifnum\@tempcnta>\@tempcntb\else\@citea
  \def\@citea{,\penalty\@m}%
  \ifnum\@tempcnta=\@tempcntb\the\@tempcnta\else
   {\advance\@tempcnta\@ne\ifnum\@tempcnta=\@tempcntb \else
\def\@citea{--}\fi
    \advance\@tempcnta\m@ne\the\@tempcnta\@citea\the\@tempcntb}\fi\fi}

\catcode`\@=12

\begin{document}

\title[Melting the CGC  at the LHC]{Melting the Color Glass Condensate at the LHC}

\author{H. Fujii${}^1$, F. Gelis${}^2$, A. Stasto${}^3$, R. Venugopalan${}^4$}

\address{{\bf 1.} Institute of Physics, University of Tokyo, Komaba, Tokyo 153-8902, Japan
}
\address{{\bf 2.} Theory Division, PH-TH, Case C01600, CERN, CH-1211 Geneva 23, Switzerland
}
\address{{\bf 3.} Physics Department, Penn State University,  
PA 16802-6300, USA
}
\address{{\bf 4.} Physics Department,
  Brookhaven National Laboratory, Upton, NY 11973, USA
}
\begin{abstract}
  The charged particle multiplicity in central
  AA collisions  and the production of heavy flavors
  in pA collisions at the LHC is predicted in the CGC framework.
\end{abstract}

\section{Introduction}
In the  Color Glass Condensate (CGC) framework,  fast (large $x$) partons are described as frozen light cone color sources while the soft
(small $x$) partons are described as gauge fields. The distribution of
the fast color sources and their evolution with rapidity is described by the JIMWLK evolution
equation; it is well approximated for large nuclei by the
Balitsky-Kovchegov (BK) equation. When two hadrons collide,  a time dependent color field is 
produced  that eventually decays into gluons~\cite{zakopane}.  When the projectile is dilute (e.g.,AA collisions at forward rapidity or  pA collisions), $k_\perp$ factorization holds for gluon production, thereby simplifying computations. For quark production, $k_\perp$ factorization breaks down and is recovered only for large invariant masses and momenta.

\section{Particle multiplicity in central AA collisions}
The $k_\perp$ factorized cross-sections are convolutions over ``dipole" scattering amplitudes in the projectile and target. Initial conditions for the BK evolution of these are specified at an initial $x=x_0$ (chosen here to be  $x_0\approx 10^{-2}$). In this work \cite{limitfrag}, we consider two  initial conditions, based respectively on the McLerran-Venugopalan (MV) model or on the Golec-Biernat--Wusthoff (GBW) model.We adjust the free parameters  to reproduce the limiting fragmentation curves measured at RHIC from$\sqrt{s}=20$~GeV to $\sqrt{s}=200$~GeV. The value of $\alpha_s$ in the {\sl fixed coupling} BK equation is tuned to obtain the observed rate of growth of the saturation scale.The rapidity distribution $dN/dy$ is converted into the
pseudo-rapidity distribution $dN/d\eta$ by asuming the produced particles have $m\sim
200$~MeV.
\begin{figure}[htbp]
\begin{center}
\resizebox*{5cm}{!}{\includegraphics{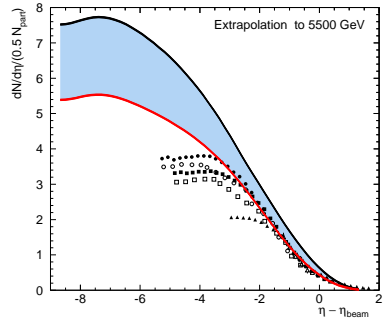}}
\end{center}
\caption{\label{fig:dndeta}Number of charged particles per unit of
  pseudo-rapidity at the LHC energy.}
\end{figure}
A prediction for AA collisions at the LHC is obtained by
changing $\sqrt{s}$ to $5.5$~GeV. From  Fig. \ref{fig:dndeta}, we can infer $dN_{\rm
  ch}/d\eta\big|_{\eta=0}=1000-1400$; the two endpoints correspond to GBW and MV initial conditions respectively.

\section{Heavy quark production in pA collisions}
\begin{figure}[htbp]
\begin{center}
\resizebox*{5cm}{!}{\includegraphics{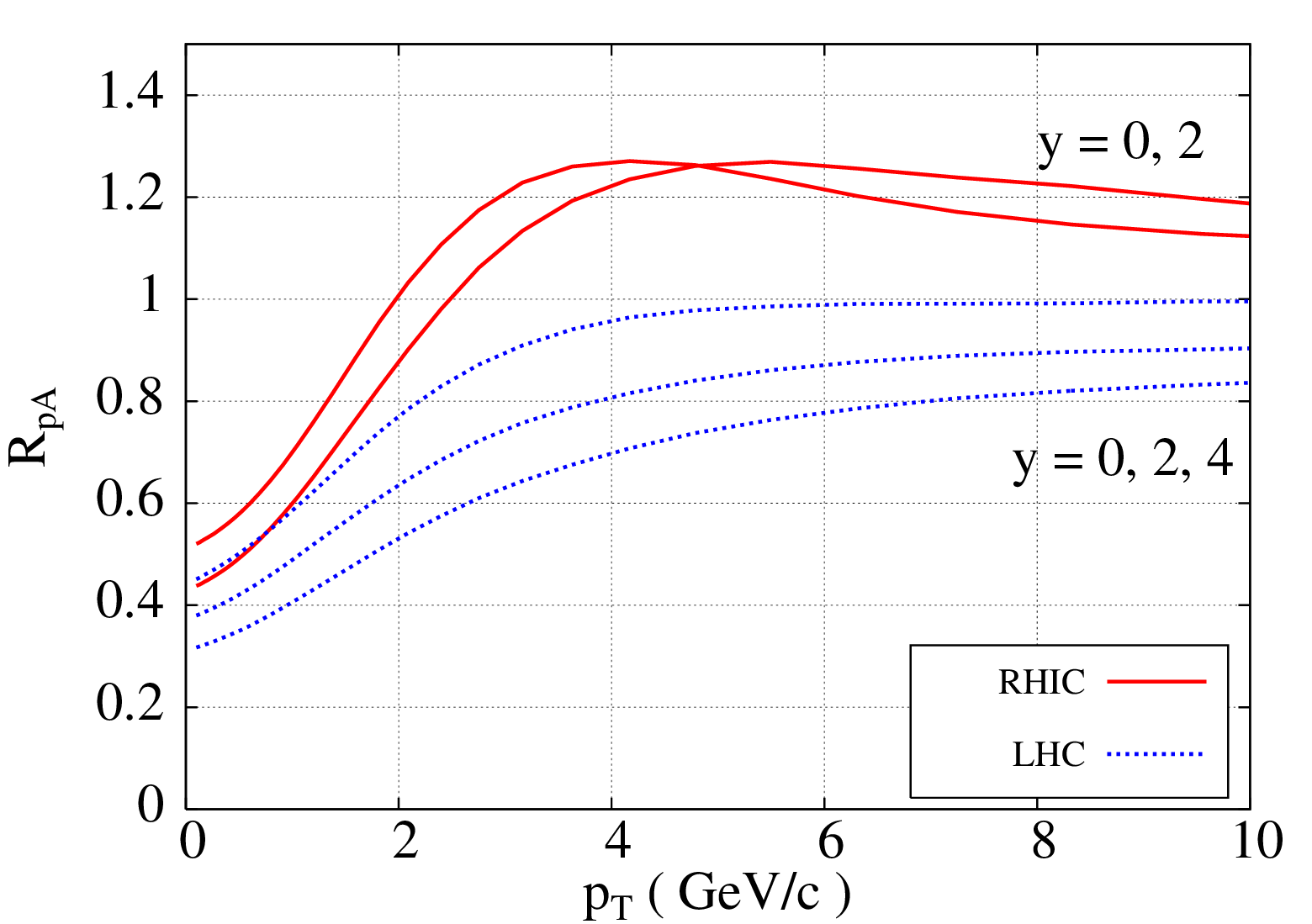}}
\hskip 5mm
\resizebox*{5cm}{!}{\includegraphics{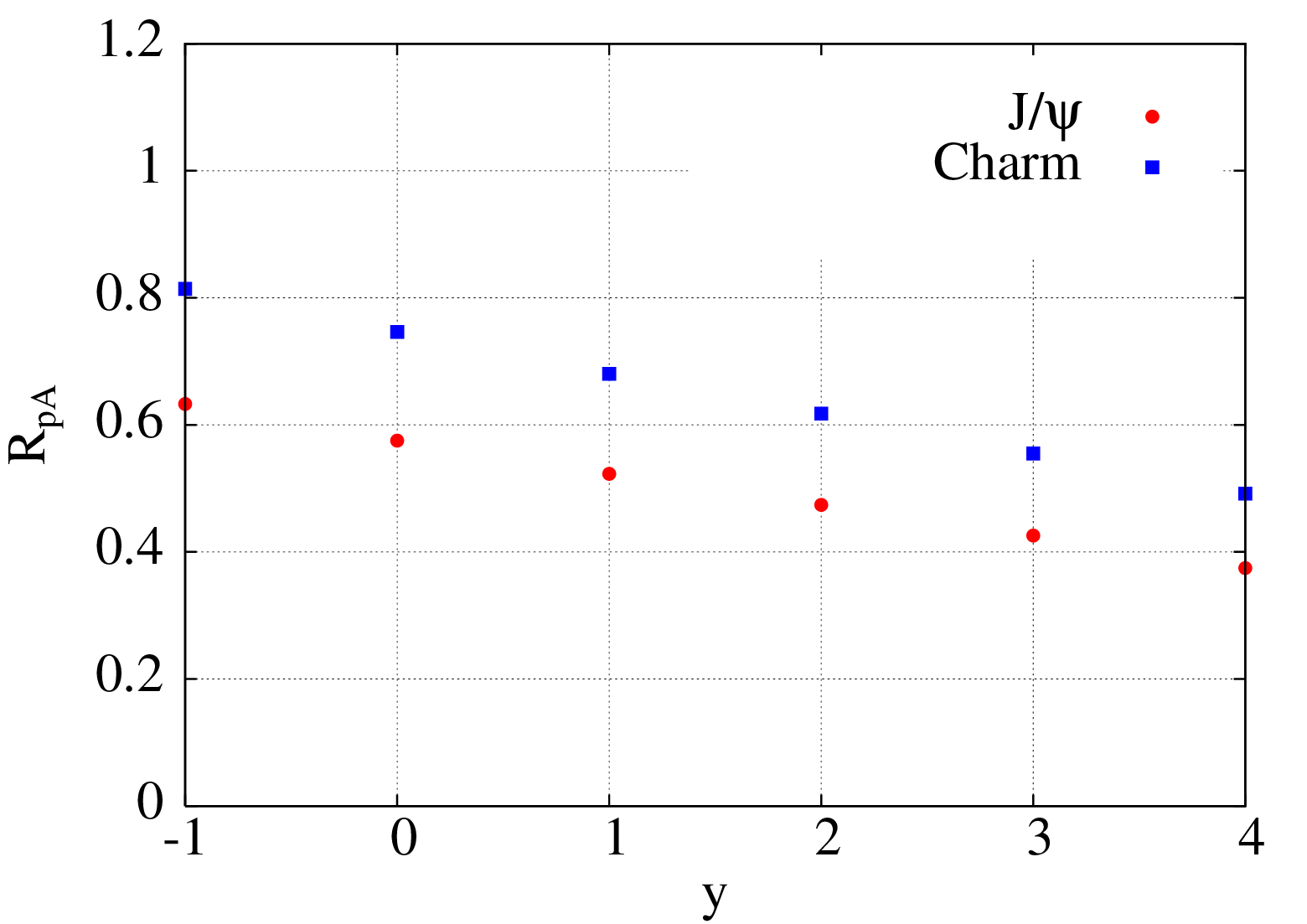}}
\end{center}
\caption{\label{fig:rpa}Left: nuclear modification factor for $D$
  mesons as a function of $p_\perp$. Right: the same ratio as a
  function of rapidity, for $D$ mesons and for $J/\psi$.}
\end{figure}

The cross-section for the
production of a pair of heavy quarks \cite{heavyquarks1} is the simplest process for which  $k_\perp$-factorization breaks down\cite{heavyquarks2} in pA collisions. This is due to the sensitivity of the cross-section to 3- and 4-point correlations in the
nucleus. Integrating out the antiquark and convoluting with a
fragmentation function, one obtains the cross-section for open heavy
flavor production, e.g., $D$ mesons. Alternatively, one can use the
Color Evaporation Model to obtain the cross-section for
quarkonia bound states.  The
nuclear modification ratio is displayed in figure \ref{fig:rpa}. The main
difference at the LHC compared to RHIC energy is that this ratio is smaller than unity already at mid rapidity, and decreases further towards the proton fragmentation region~\cite{heavyquarks3}.

\end{document}